\documentclass[twocolumn,showpacs,amsmath,amssymb,aps,10pt,reprint]{revtex4-1}
\usepackage{graphics}
\usepackage{amsmath}
\usepackage{amssymb}
\usepackage[breaklinks=true,colorlinks=true,linkcolor=blue,urlcolor=blue,citecolor=blue]{hyperref}
\usepackage{soul,color}
\usepackage{epstopdf}
\usepackage{balance}
\usepackage{multirow}
\usepackage{float}
\usepackage{rotating}

\begin{document}

\title{\textbf{\Large Pinning effects on hot-electron vortex flow instability in superconducting films}}

\author{Valerij~A.~Shklovskij}

\address{Physics Department, V. Karazin Kharkiv National University, 61022 Kharkiv, Ukraine}

\date{\today}

\begin{abstract}
The hot-electron vortex flow instability in superconducting films in magnetic field $B$ at substrate temperature $T_0 \ll T_c$ is theoretically considered in the presence of pinning. The magnetic field dependences of the instability critical parameters (electric field $E^\ast$, current density $j^\ast$, resistivity $\rho^\ast$, power density $P^\ast$ and vortex velocity $v^\ast$) are derived for a cosine and a saw-tooth washboard pinning potential and compared with the results obtained earlier by M.\,Kunchur [\href {\doibase 10.1103/PhysRevLett.89.137005}{Phys. Rev. Lett. \textbf{89} (2002) 137005}] in absence of pinning. It is shown that the $B$-behavior of  $E^\ast$, $j^\ast$ and $\rho^\ast$ is \emph{monotonic}, whereas the $B$-dependence of $v^\ast$ is quite different, namely $dv^\ast/dB$ may \emph{change its sign twice}, as sometimes observed in experiments. The simplest heat balance equation for electrons in low-$T_c$ superconducting films is considered within the framework of the two-fluid model. A theoretical analysis reveals that the instability critical temperature $T^\ast \approx 5T_c/6$ at $T_0 < T^\ast/2$ with $T^\ast$ being independent of $B$.
\end{abstract}

\maketitle
\section{Introduction}
It is well known that vortex motion under the action of an applied current in superconducting films in a perpendicular magnetic field $B$ at high dissipation levels becomes unstable at some critical vortex velocity $v^\ast$. For the flux flow regime at temperatures near the superconducting transition temperature $T\lesssim T_c$ this instability was theoretically treated by Larkin and Ovchinnikov~(LO)~\cite{Lar75etp}. Their theory predicts that $v^\ast$ is \emph{independent} of $B$ that was experimentally confirmed for low-$T_c$ \cite{Kle85ltp,Vol92fnt,Per05prb} and high-$T_c$ \cite{Doe94prl,Xia96prb,Xia99prb} superconducting films. In subsequent experiments, a crossover from the magnetic-field independent behavior at high fields to $v^\ast\propto B^{-1/2}$ at low fields was reported by Doettinger \emph{et al.} \cite{Doe95pcs}. This low-field behavior was explained \cite{Doe95pcs} as $v^\ast$ multiplied with the inelastic quasiparticle scattering time must reach at least the intervortex distance to ensure the spatial homogeneity of the nonequilibrium quasiparticle distribution the LO theory is relying upon.

However, experiments performed on YBCO films at low temperatures $T\ll T_c$~\cite{Kun02prl,Kni06prb} showed an instability with a universal dependence $v^\ast\propto B^{-1/2}$ whose underlying physical mechanism was essentially different from the LO instability picture. With an account for the Bardeen-Stephen nonlinear conductivity \cite{Bar65prv}, Kunchur has shown \cite{Kun02prl,Kni06prb} that this new behavior can be explained by a simple model in which the electron gas has a thermal-like Fermi distribution function characterized by a higher temperature than that of the phonons and the bath. In contradistinction with the standard LO picture, the main effects in the Kunchur instability~\cite{Kun02prl,Kni06prb} are a rise of the electronic temperature, creation of additional quasiparticles, and a diminish of the superconducting gap. The vortex expands rather than shrinks, and the viscous drag is reduced because of a softening of gradients of the vortex profile rather than a removal of quasiparticles from the vortex core, as supposed within the framework of the LO theory. While the electron temperature rises, the resulting resistivity increase leads to a decrease in current above a certain value of the electric field. That is, the current-voltage-curve (CVC) becomes non-monotonic in $j$ and exhibits an electric field instability. All experimental observables for the hot-electron instability were calculated in Ref. \cite{Kun02prl}. The experimental results on YBCO were successfully fitted to the predicted $B$-dependences and $j(E)$ curves \emph{in absence of pinning} without any adjustable parameters~\cite{Kun02prl,Kni06prb}.

The objective of this paper is to theoretically consider the hot-electron vortex flow instability in low-$T_c$ superconducting thin films at $T \ll T_c$ in the \emph{presence of pinning}. This study is motivated by two aspects. Firstly, low-$T_c$ superconductors are characterized by a simple electronic structure, thus allowing one to use a more simple heat balance equation than that for YBCO. Secondly, vortex pinning in low-$T_c$ films is usually stronger than in high-$T_c$ epitaxial films so that it has to be properly taken into account. It should be emphasized that neither LO \cite{Lar75etp} nor Kunchur \cite{Kun02prl} approaches capture vortex pinning in the physical picture of the flux-flow instability in the nonlinear CVC. In experimental samples, however, vortex pinning is omnipresent and there is growing interest in addressing pinning effects on the instability critical parameters in superconductors
\cite{Leo10pcs,Gri12apl,Gri15prb,Leo16prb}, in particular those with artificial pinning structures \cite{Sil12njp}. While a recent theoretical account for the LO instability at $T \simeq T_c$ can be found in Ref. \cite{Shk17snd}, the respective generalization of the Kunchur approach at temperatures $T\ll T_c$ has not been elaborated so far.

Both these aspects will be addressed in this paper. Namely, Sec. \ref{SecInst} presents a phenomenological approach to account for pinning effects on two simplest CVCs in the flux-flow regime. These CVCs are exemplary, as they are calculated at $T=0$ for two pinning potentials of the washboard type, namely for a cosine washboard pinning potential (WPP) \cite{Shk08prb,Shk11prb} and for a saw-tooth WPP~\cite{Shk99etp,Shk06prb}. A cosine WPP is widely used in theoretical papers, see e.g. Refs. \cite{Mar76prl,Che91prb,Cof91prl,Maw97prb,Maw99prb,Shk14pcm}. At the same time, both model WPPs are realistic as they can be realized by various experimental techniques, see e.\,g. Ref. \cite{Dob17pcs} for a review. For instance, both WPPs can be used for modelling the resistive responses of nanopatterned superconductors with uniaxial anisotropic pinning induced either by ferromagnetic stripes deposited onto the film surface \cite{Dob10sst,Dob11pcs,Dob11snm} or nanogrooves milled in the film \cite{Dob11snm,Dob12njp,Dob16sst}. In addition, the understanding of pinning effects on the flux-flow instability is the key to expanding the current-operation range of microwave applications \cite{Lar15nsr,Dob15apl,Dob15met,Sil17inb} and it is crucial for the development of superconducting devices of the fluxonic type \cite{Dob17pcs}. Both WPPs allow one to reproduce the calculation of the hot-electron instability in the spirit of Refs. \cite{Kun02prl,Kni06prb} and to solve a more simple heat balance equation within the framework of the two-fluid model in Sec. \ref{SecPow}. While in the limiting case of no pinning the results of Refs. \cite{Kun02prl,Kni06prb} are recovered, the presented in what follows approach provides a more simple and intuitively clearer physics.

\section{Instability parameters}
\label{SecInst}
\subsection{Problem definition}
\begin{figure}
\centering
\includegraphics[width=1\linewidth]{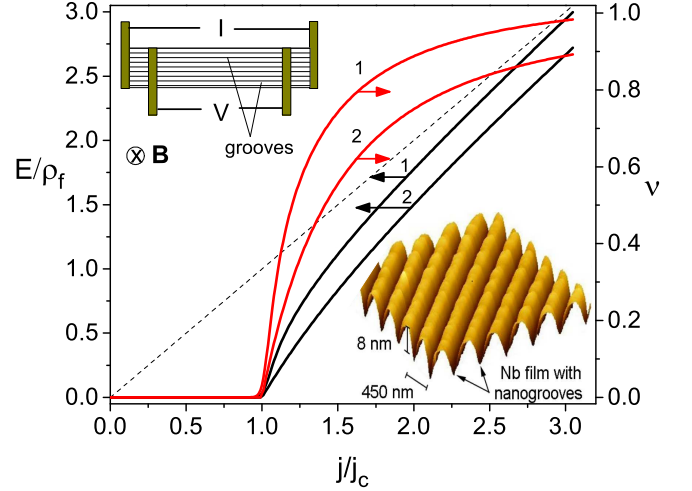}
\caption{Left axis: The nonlinear current-voltage curve $E(j)$ (red online) calculated in the limit of low temperatures for a cosine (blue online) WPP of Refs. \cite{Shk08prb,Shk11prb} (\emph{curve 1}) and a saw-tooth (black online) WPP of Refs. \cite{Shk99etp,Shk06prb} (\emph{curve 2}). The dashed line corresponds to the free flux-flow regime $E/j = \rho_f$, where $\rho_f$ is the flux-flow resistivity. Right axis: The respective nonlinear functions $\nu(j)$ calculated by Eq. (27) of Ref. \cite{Shk08prb}  (\emph{curve 1}) and Eq. (28) at $\epsilon=1$ of Ref. \cite{Shk99etp} (\emph{curve 2}). Inset: Atomic force microscope image of a Nb film surface with a nanogroove array milled by focused ion beam \cite{Dob16sst} and inducing a pinning potential of the washboard type.}
\label{fig1}
\end{figure}

The effect of a WPP on the flux-flow overheating instability is considered at substrate temperature $T_0\ll T_c$, as earlier studied by Kunchur in absence of pinning \cite{Kun02prl}. For simplicity, the problem is considered at $T_0 = 0$, when the transport current flows along the WPP channels, refer to the upper inset in Fig.\,\ref{fig1}. In this geometry the vortices experience the action of the Lorentz force in the direction transverse to the pinning channels. The respective nonlinear CVC of the sample can be presented as
\begin{equation}
\label{eCVC}
            \sigma E = j\nu(j).
\end{equation}
Here $E$ is the longitudinal electric field, $j$ is the density of the transport current, and $0\leq \nu(j) \leq 1$ is a nonlinear function with the condition $\nu(j) = 0$ for $j < j_c$, where $j_c$ is the critical (depinning) current density, refer to Fig. \ref{fig1}. The nonlinear function $\nu(j)$ appears in Eq. (\ref{eCVC}) due to the effect of the WPP on the vortex dynamics. In Eq.~\eqref{eCVC} $\sigma = \sigma(T) = \sigma_nH_{c2}(T)/B$ is the temperature-dependent Bardeen-Stephen~\cite{Bar65prv} flux-flow conductivity, $\sigma_n$ is the normal metal film conductivity at $T\approx T_c$, $H_{c2}$ is the upper critical field, and $B$ is the flux density applied perpendicular to the film.

If $j_c\rightarrow 0$, then $\nu(j) \rightarrow 1$ and the linear CVC $\sigma E = j$ follows from Eq.~\eqref{eCVC}. The expression $\sigma E = j$ was used by Kunchur \cite{Kun02prl,Kni06prb} as the initial form of the CVC. Two different WPP forms resulting in two different $\nu(j)$ functions plotted in Fig. \ref{fig1} will be considered next.

\subsection{Cosine pinning potential}
\label{ssCosine}
For the cosine WPP, $\nu(j) = \sqrt{1 - (j_c/j)^2}$~\cite{Shk08prb} and
\begin{equation}
\label{eCosine}
    \sigma E = \sqrt{j^2 - j_c^2} \qquad \mathrm  {or}   \qquad  j = \sqrt{j_c^2 + \sigma ^2 E^2}.
\end{equation}
In the overheating approach of Kunchur \cite{Kun02prl,Kni06prb}, in the vortex state of a film with quasiparticles temperature $T = T(E)$ the CVC instability in Eq.~\eqref{eCosine} appears as a region of negative differential conductivity, where $j$ decreases as a function of $E$. The values of the instability points $j^\ast$ and $E^\ast$ can be determined from a set of equations which include the heat balance equation
\begin{equation}
\label{eHeatBalance}
    P\tau_e = \int_0^T C(T^\prime)dT^\prime
\end{equation}
and the CVC extremum condition
\begin{equation}
\label{eCVCextremum}
   \frac{dj}{dE}\Big|_{E=E^\ast} = 0.
\end{equation}
Here $P = jE$ is the dissipated power, $\tau_e(T)$ is the energy relaxation time, and $C(T)$ is the electronic specific heat per unit volume. As follows from Eq.~\eqref{eCVCextremum},
\begin{equation}
\label{eCondition}
   \frac{dE}{dT}\Big|_{E=E^\ast} = -E^\ast[\sigma^\prime(T)/\sigma(T)]\Big|_{T=T^\ast},
\end{equation}
where the prime denotes differentiation with respect to temperature. Substitution of Eq.~\eqref{eCondition} into the relation $d(P\tau_e)/dT = C(T)$, following form Eq.~\eqref{eHeatBalance}, leads to the expression
\begin{equation}
\begin{array}{lll}
\label{eCsigma}
   C(T^\ast)\sigma(T^\ast) = E^\ast [\tau^\prime(T^\ast)\sigma(T^\ast) - \tau(T^\ast)\sigma^\prime(T^\ast)]\times\\[2mm]
   \qquad\qquad\qquad\qquad\qquad\qquad\times\sqrt{j_c^2 + \sigma^2(T^\ast) E^{\ast 2}},
\end{array}
\end{equation}
which in absence of pinning (i.\,e. when $j_c = 0$) reduces to the expression for $E_0^\ast$ (see also Eq. (5) in Ref. \cite{Kni06prb}).
\begin{equation}
\label{eE0ast}
    E_0^{\ast 2} = C(T^\ast)\rho_n B/[H_{c2}(T)\tau^\prime_e(T) + H_{c2}^\prime(T)\tau_e(T)]\Big|_{T = T^\ast}.
\end{equation}
Taking the square of Eq.~\eqref{eCsigma} it is easy to show that $z \equiv (E^\ast/E_0^\ast)^2$ can be found from the equation
\begin{equation}
\label{eZ2}
    z^2 + 2\mu z - 1 = 0,
\end{equation}
where the dimensionless parameter $\mu$ links the instability problem with and without pinning through the relation
\begin{equation}
\label{e2mu}
    2\mu \equiv (j_c/j_0^\ast)^2.
\end{equation}
For $j_c = 0$ one has $\mu = 0$ and $z =1$, i.\,e. one returns to the problem discussed in Refs.~\cite{Kun02prl,Kni06prb}. In the general case $0 \leq \mu <\infty$, and the solution of Eq.~\eqref{eZ2} reads
\begin{equation}
\label{eZsolution}
    z = \sqrt{1 + \mu^2} - \mu = 1/(\sqrt{1 + \mu^2} + \mu).
\end{equation}
From Eq.~\eqref{eZsolution} it follows that $z(\mu)$ monotonically decreases with increasing $\mu$, i.\,e. $E^\ast$ decreases with increasing $j_c$. Next, from Eq.~\eqref{eCosine} it follows $j^{\ast2} = j_c^2 + z j_0^{\ast2}$ and, if we define $y\equiv(j^\ast/j_0^\ast)^2$,
\begin{equation}
\label{eY}
    y = 1/z = \sqrt{1 + \mu^2} + \mu.
\end{equation}
From Eq.~\eqref{eY} it follows that $y(\mu)$ monotonically increases, i.\,e. $j^\ast$ increases with increasing $j_c$.

Now, having analyzed the $\mu$-behavior of $E^\ast$ and $j^\ast$, it is possible to derive the $\mu$-dependences of several related responses at the instability point. These responses are the critical vortex resistivity $\rho^\ast = E^\ast/j^\ast$, the critical vortex velocity $v^\ast/c = E^\ast/B$, and the dissipated power $P^\ast = E^\ast j^\ast$. Accordingly, using Eqs.~\eqref{eZsolution} and~\eqref{eY} one concludes that the critical velocity $v^\ast(\mu) \sim E^\ast(\mu)$ is monotonically decreasing in $\mu$, while $P^\ast = P^\ast_0$  does not depend on $\mu$, and
\begin{equation}
\label{eRast}
    \rho^\ast = \rho^\ast_0/(\sqrt{1 + \mu^2} + \mu).
\end{equation}
From Eq.~\eqref{eRast} it follows that $\rho^\ast(\mu)$ monotonically decreases in $\mu$, i.\,e. $\rho^\ast$ decreases as $j_c$ increases.

After the analysis of the $j_c$-behavior of the critical parameters $E^\ast, j^\ast, \rho^\ast, P^\ast$, and $v^\ast$, it is instructive to analyze also their $B$-dependences at $j_c = const$. In other words, for a moment it is supposed that $j_c$ is independent of $B$. To proceed with the $B$-analysis, it is necessary to remind the $B$-dependences of the critical parameters $E^\ast_0, j^\ast_0, \rho^\ast_0, P^\ast_0$, and $v^\ast_0$ in absence of pinning (i.\,e. at $j_c =0$). Previously it was shown \cite{Kun02prl,Kni06prb} that $E_0^\ast = \kappa\sqrt{B}$, $j_0^\ast = \gamma/\sqrt{B}$, $\rho_0^\ast = \alpha B$, $v_0^\ast/c = \kappa/\sqrt{B}$, and $P_0^\ast$ is independent of $B$. Here the two constants $\alpha$ and $\gamma$ have been introduced such that $\kappa =\alpha\gamma$. Their values can be obtained from theory and compared with experiment, see Refs. \cite{Kun02prl,Kni06prb}. Then it follows that $\mu = j_c^2/2j_0^{\ast2} = \varepsilon_c B$, where $\varepsilon_c = j_c^2/2\gamma^2$. From the latter it follows that $\mu$ is an increasing function of $j_c$ and $B$.

Unfortunately, a direct inspection is not sufficient to check Eqs.~\eqref{eZsolution}--\eqref{eRast} for monotonicity in $B$. The corresponding critical parameters and their $B$-derivatives should be calculated for this. For $dE^\ast/dB$ one has
\begin{equation}
\label{edEdB}
        dE^\ast/dB = \kappa/2\sqrt{B}\sqrt{1+\mu^2}(\sqrt{1+\mu^2} +\mu)^{3/2} >0.
\end{equation}
As it follows from Eq.~\eqref{edEdB}, $E^\ast(B)$ monotonically increases with growing $B$ while $dE^\ast/dB$ decreases. The behavior of $\rho^\ast(B)$ is similar, because
\begin{equation}
\label{edRdB}
        d\rho^\ast/dB = \alpha/\sqrt{1+\mu^2}(\sqrt{1+\mu^2} +\mu)^2 >0,
\end{equation}
 i.\,e. it monotonically increases with growing $B$ while $d\rho^\ast/dB$ decreases. For $dj^\ast/dB$ one obtains
\begin{equation}
\label{edJdB}
        dj^\ast/dB = -\gamma/2B^{3/2}\sqrt{1+\mu^2}(\sqrt{1+\mu^2} +\mu) <0.
\end{equation}
From Eq. \eqref{edJdB} it follows that $j^\ast(B)$ monotonically decreases with growing $B$ while $dj^\ast/dB$ decreases. Finally, it is interesting to derive the $B$-dependence of the critical vortex velocity $v^\ast/c = E^\ast/B$, which for the LO instability~\cite{Lar75etp} does not depend on $B$ (see also the Bezuglyj-Shklovskij (BS) generalization of the LO instability where the $B$-dependence appears for fields larger than the overheating field $B>B_T$ \cite{Bez92pcs}). It can be shown that
\begin{equation}
\label{edVdB}
        dv^\ast/dB = -(c\kappa/2B^{3/2})\sqrt{\sqrt{1+\mu^2}+\mu}/\sqrt{1+\mu^2} <0.
\end{equation}
From Eq. \eqref{edVdB} it follows that $v^\ast(B)$ monotonically decreases with growing $B$ and $d v^\ast/dB$ does so.

Up to this point, the analysis of Eqs.~\eqref{edEdB}-\eqref{edVdB} was done for $j_c$ being $B$-independent, when $\mu = j_c^2/j_0^{\ast2} = \varepsilon_c B$ was proportional to $B$. In reality, however, $j_c$ depends upon $B$ and $\mu(B) = j_c^2(B)B/2\gamma^2$ has a more complex $B$-dependence. In order to analyze the $B$-dependence of $\mu(B)$ following from the $j_c(B)$ behavior, the following scaling is assumed for simplicity
\begin{equation}
\label{eScaling}
        j_c(B) = j_B(B_c/B)^m,
\end{equation}
where $j_B$ and $B_c$ are fitting parameters which provide correct values of $j_c(B)$, while the exponent $m>0$ is the main parameter which determines the $B$-behavior of $\mu(B)$. It is clear from Eq.~\eqref{eScaling} that for $m=0$ one returns to the $B$-independent case $j_c = const$, while for $m=1/2$ one has $\mu(B)=const$, i.\,e. it is independent of $B$. Hence, for the determination of the $B$-dependence of the critical parameters it is necessary to calculate the derivative $d\mu(B)/dB$. Whereas for $j_c=const$ the derivative $d\mu/dB = \varepsilon_c$ was $B$-independent, now it reads
\begin{equation}
\label{edMdB}
        d\mu(B)/dB = (1-2m)\mu(B)/B.
\end{equation}
From Eq. \eqref{edMdB} it follows that $d\mu/dB$ equals to zero and changes its sign at $m=1/2$. In other words, $\mu(B)$ \emph{decreases} with growing $B$ for $m>1/2$, whereas $\mu(B)$ \emph{increases} for $0<m<1/2$.

Now it is time to turn to an analysis of the influence of the dependences given by Eqs. \eqref{eScaling} and \eqref{edMdB} on the $B$-dependence of the critical parameters $E^\ast, j^\ast, \rho^\ast, v^\ast$ and their $B$-derivatives. Since
\begin{equation}
\label{eEvB}
        E^\ast(B) = \kappa\sqrt{B}/\sqrt{\sqrt{1+\mu^2}+\mu},
\end{equation}
then it follows that the denominator in Eq.~\eqref{eEvB} is decreasing with growing $B$ for $m>1/2$, thereby resulting in $E^\ast(B)$ increasing more rapidly than $E^\ast_0 = \kappa\sqrt{B}$. For $m<1/2$ the denominator is increasing with growing $B$ and, hence, the derivative $dE^\ast/dB$ should be analyzed. The result is
\begin{equation}
\label{edEvB}
        \displaystyle\frac{dE^\ast}{dB} = \displaystyle\frac{\kappa}{2\sqrt{B}}\frac{1+2m\mu(\sqrt{1+\mu^2}+\mu)}{\sqrt{1+\mu^2}(\sqrt{1+\mu^2}+\mu)^{3/2}}>0.
\end{equation}
As follows from Eq. \eqref{edEvB}, $E^\ast$(B) increases with growing $B$ for any $m>0$, but the rate of this increase depends upon whether $m>1/2$ or $0<m<1/2$. It is instructive to point out that the $\mu(B)$ dependence reads
\begin{equation}
\label{eMuvB}
        \mu(B) = j_c^2(B)B/2\gamma^2 = KB^{1-2m},
\end{equation}
where $K = j_B^2B_c^2/2\gamma^2$. Equation \eqref{edMdB} follows at once from Eq. \eqref{eMuvB}.

As for $j^\ast(B)$, one has [see Eq. \eqref{eY}]
\begin{equation}
\label{ejAst}
        j^\ast = j_0^\ast\sqrt{\sqrt{1+\mu^2}+\mu} = (\gamma/\sqrt{B})\sqrt{\sqrt{1+\mu^2}+\mu}.
\end{equation}
It follows from Eq.~\eqref{ejAst} that for $m>1/2$, $j^\ast(B)$ is decreasing faster with growing $B$ than $j_0^\ast(B) = \gamma/\sqrt{B}$, whereas for $m<1/2$ the situation is not clear because of the $\mu$-dependent multiplier in Eq. \eqref{ejAst} increasing with growing $B$. The calculation of $dj^\ast/dB$ yields
\begin{equation}
\label{edjAstdB}
        \displaystyle\frac{dj^\ast}{dB} = -\displaystyle\frac{\gamma}{2B\sqrt{B}}\sqrt{\sqrt{1+\mu^2}+\mu}\left[1-\displaystyle\frac{(1-2m)\mu}{\sqrt{1+\mu^2}}\right] <0,
\end{equation}
because at any $m>0$ and $\mu$ the expression in the brackets is positive.

Now it is possible to write down an expression for the $B$-dependent the resistivity at the instability point
\begin{equation}
\label{eRfvB}
        \rho^\ast(B)=\alpha B/(\sqrt{1+\mu^2}+\mu).
\end{equation}
From Eq.~\eqref{eRfvB} it follows that $\rho^\ast(B)$ is increasing with growing $B$ more rapidly than $\rho^\ast_0(B) = \alpha B$ for $m > 1/2$ due to the denominator of Eq.~\eqref{eRfvB} decreasing with growing $B$. For $m < 1/2$, again, the derivative $d\rho^\ast/dB$ should be calculated. This yields
\begin{equation}
\label{edRAstdB}
        d\rho^\ast/dB = \alpha \left[1 -\displaystyle\frac{(1-2m)\mu}{\sqrt{1+\mu^2}}\right]/(\sqrt{1+\mu^2} + \mu) >0
\end{equation}
because as in Eq.~\eqref{edjAstdB} the expression in the brackets is positive, i.\,e. $\rho^\ast(B)$ always increases with growing $B$. Finally, the $B$-dependence of $v^\ast$ should be considered
\begin{equation}
\label{eVastB}
        v^\ast(B) = c\kappa /\sqrt{B}\sqrt{\sqrt{1+\mu^2} + \mu}.
\end{equation}
From Eq.~\eqref{eVastB} it follows that for $m< 1/2$ $v^\ast(B)$ is decreasing with growing $B$ faster than $v^\ast_0(B) = c\kappa/\sqrt{B}$ and for $m> 1/2$ the derivative $dv^\ast/dB$ should be calculated. The result is
\begin{equation}
\label{edVAstdB}
        dv^\ast/dB = -c\kappa \left[1 +\displaystyle\frac{(1-2m)\mu}{\sqrt{1+\mu^2}}\right]/2B\sqrt{B}\sqrt{\sqrt{1+\mu^2} + \mu}.
\end{equation}
Equation~\eqref{edVAstdB} reduces to Eq.~\eqref{edVdB} in the limit $m=0$, when $dv^\ast/dB<0$. For $m>1/2$ it is easy to show that the bracket in Eq.~\eqref{edVAstdB} may be negative at $m>(1 + \sqrt{1+1/\mu^2})/2$. In consequence of this one has $m\simeq 1 + 1/(2\mu)^2$ for $\mu\gtrsim 2$. In this case $dv^\ast/dB > 0$, i.\,e. it \emph{changes its sign} when $B\rightarrow 0$.

The new results given by Eqs. \eqref{edEvB}-\eqref{edVAstdB} derived using the $j_c(B)$ and $\mu(B)$ dependences given by Eqs.~\eqref{eScaling} and~\eqref{eMuvB}, respectively, can be briefly summarized as follows. The main result for the $B$-dependences of the critical parameters $E^\ast(B)$, $j^\ast(B)$, $\rho^\ast(B)$ and $P^\ast(B)$ consists in \emph{maintaining the monotonicity of their $B$-dependences} for the case $j_c = j_c(B)$ given by Eq.~\eqref{eScaling}. In other words, the $B$-derivatives of these parameters maintain the same sign as for $j_c = const$, see Eqs. \eqref{edEvB}, \eqref{edjAstdB}, and \eqref{edRAstdB}. At the same time, for the $B$-dependent critical current given by Eq.~\eqref{eScaling} a \emph{sign change} of $dv^\ast/dB$ is possible for $m\gtrsim1$, see Eq.~\eqref{edVAstdB} and the subsequent discussion. That is, the monotonicity of $v^\ast(B)$ at small $B$ may be \emph{violated}. Moreover, since usually $j_c(B)$ at small $B$ can be approximated again by Eq.~\eqref{eScaling} with $m<1/2$, then there may be a \emph{second sign change} in $dv^\ast/dB$ at $B$ close to the first critical field $B_{c1}(T)$. This is sometimes observed in experiments \cite{Gri09pcm,Leo10pcs,Gri10prb,Gri12apl,Gri11snm,Sil12njp}. To conclude, the presented here phenomenological approach using the experimentally measured $B$-dependent critical current provides a simple physics which can explain the nonmonotonic behavior of $v^\ast(B)$ at small $B$.

\subsection{Saw-tooth pinning potential}
\label{SubsectsST}
For the saw-tooth WPP \cite{Shk99etp}, $\nu(j) = 1 - (j_c/j)^2$ for $j > j_c$ and $\nu(j) = 0$ for $0< j < j_c$. Then, $\sigma E = (j^2 - j_c^2)/j$ for $j>j_c$ or
\begin{equation}
\label{eSaw}
        j = (\sigma E/2)[1 + \sqrt{1+(2j_c/\sigma E)^2}]
\end{equation}

Repeating the steps, which were detailed above for the cosine WPP in Sec.~\ref{ssCosine}, it can be shown that
\begin{equation}
\label{eEE0}
        E^\ast/E_0^\ast = 1/\sqrt{1+x},
\end{equation}
where $x = 2\mu = (j_c/j_0^\ast)^2$ [see also Eq.~\eqref{e2mu}]. Then
\begin{equation}
\label{ejj0}
        j^\ast/j_0^\ast = \sqrt{1+x},
\end{equation}
\begin{equation}
\label{eRR0}
        \rho^\ast/\rho_0^\ast = 1/(1+x),
\end{equation}
and, finally, $P = P^\ast$.

Qualitatively, the $x$-behavior of the critical parameters given by Eqs.~\eqref{eEE0}-\eqref{eRR0} is similar to the $\mu$-behavior of the analogous quantities for the cosine WPP, see Eqs.~\eqref{eZsolution}-\eqref{eRast}. This similarity can also be extended to the $B$-behavior of the $B$-derivatives of $E^\ast$, $j^\ast$, $\rho^\ast$, and $v^\ast$ given previously by Eqs.~\eqref{edEdB}--\eqref{edVdB} for $j_c = const$. Moreover, using the $B$-dependent critical current $j_c(B)$ given by Eq.~\eqref{eScaling}, it is possible to repeat qualitatively all the conclusions about the monotonicity of $E^\ast(B)$, $j^\ast(B)$, $\rho^\ast(B)$, and the possible non-monotonicity of $v^\ast(B)$. For the latter the exact expression reads
\begin{equation}
\label{eVastdB}
        \frac{dv^\ast}{dB}= -\displaystyle\frac{c\kappa[1 + 2x(1-m)]}{2\sqrt{B}B(1+x)^{3/2}}
\end{equation}
From Eq.~\eqref{eVastdB} it follows that the bracket in the denominator in Eq.~\eqref{eVastdB} may be negative at $m>1$ and $x> 1/2(m-1)$. In this case one has $dv^\ast/dB > 0$, i.\,e. it changes its sign at $B\rightarrow 0$. Summarizing these short comments on the $j_c(B)$-behavior for the saw-tooth WPP, one can state that the main results on the monotonicity of the critical parameters behavior for both CVC types are qualitatively similar, i.\,e. a \emph{particular form of the WPP does not affect the considered physics}.

\section{Dissipated power and quasiparticles temperature}
\label{SecPow}
\subsection{Two-fluid approach}
In Kunchur's approach~\cite{Kun02prl}, the quasiparticles temperature with respect to the substrate temperature $T_0 \ll T_c$ can be determined by \emph{numerical} integration of the heat balance equation~\eqref{eHeatBalance}, where the temperature-dependent functions $\tau_e(T)$ and $C(T)$ were specifically calculated for the considered YBCO sample. As follows from Eq.~\eqref{eHeatBalance} at given $\tau_e(T)$ and $C(T)$, the quasiparticles temperature $T$ depends on the dissipated power $P = Ej$ and $T_0$. In~\cite{Kun02prl,Kni06prb} it was shown that for YBCO the quasiparticles temperature at the instability point of the CVC, $T^\ast(P)$, \emph{weakly depends} on $T_0$ and equals to approximately $76$\,K, i.\,e. $T^\ast$ is \emph{not close} to $T_c \simeq 90$\,K.

In what follows the $T^\ast(P,T_0)$ dependence will be estimated for a more simple case of a low-temperature superconductor film like Nb \cite{Dob12tsf}. In this case the same physics of quasiparticles overheating can be explained by a more simple heat balance equation than Eq.~\eqref{eHeatBalance}. The main features of this more simple approach were presented by BS in~\cite{Bez92pcs}, see sections 1 and 2 therein. In the BS approach it is supposed that $P(T,T_0)$ dependence can be approximated by the same expression, see Eq.~(18) in~\cite{Bez92pcs}, as for normal electrons at temperature $T$ near $T_c$ as it can be made within the framework of the two-fluid model of superconductivity~\cite{Tin04boo}. It will be shown that this approach yields $T^\ast$ near $T_c$ (but not too close to $T_c$ where the mechanism of the LO instability~\cite{Lar75etp} dominates).

For the heat flow $Q$ from the film to the substrate one has the equation
\begin{equation}
\label{eQ}
        Q = Ad[(kT)^5 - (kT_0)^5)],
\end{equation}
which is accurate to corrections of the order of $(\Delta/T)^2\ll 1$, where $\Delta(T)$ is the superconducting gap. In Eq.~\eqref{eQ} $d$ is the film thickness and $A$ is a coefficient which is not essential for the following reasoning and it is given by Eq.~(18) of Ref.~\cite{Bez92pcs}. Equation~\eqref{eQ} describes the case when nonequilibrium phonons escape from a thin film without reabsorption by quasiparticles. The heating regime of the film in this limit is known as electron overheating~\cite{Bez92pcs}, termed so as one describes the quasiparticles and phonons by different temperatures, $T$ and $T_0$, respectively. Taking into account that $Q = Pd$, where $P = Ej$, from Eq.~\eqref{eQ} follows
\begin{equation}
\label{eP}
        P = A[(kT)^5 - (kT_0)^5)].
\end{equation}

First, the critical parameters will be considered in this approach without pinning, i.\,e. the calculations of Kunchur~\cite{Kun02prl} will be repeated for the case when the heat balance equation~\eqref{eHeatBalance} has the form of Eq.~\eqref{eP}. Since $P = \sigma(T)E^2$, where $\sigma(T) = \sigma_nH_{c2}(T)/B$ and $T$ is supposed to be close to $T_c$, it is possible to write $H_{c2}(T) \simeq Rk(T_c-T)$, where $R = 4c/\pi e D$ is valid for superconductors with a short mean free path and diffusivity $D$ ~\cite{Lar75etp}. Then, from Eq.~\eqref{eP} it follows that
\begin{equation}
\label{eE2T}
        E^2(T) = Z^2 B[(kT)^5 - (kT_0)^5)]/k(T_c - T),
\end{equation}
where $Z^2 = A/R\sigma_n$. From Eq.~\eqref{eE2T} it follows that for $T_0 < T/2$ one may neglect $T_0$. If also $\theta \equiv T_c - T$ is rather small, i.\,e. $\theta\ll T_c$, then  in Eq.~\eqref{eE2T} it is possible to change $T\rightarrow T_c$ in the bracket because $(kT)^5\simeq(kT_c)^5(1- 5\theta/T_c)$ and in this limit the main $T$-dependence of $E^2(T)$ on $\theta$ is
\begin{equation}
\label{eE2Tlimit}
        E^2(T)\simeq Z^2 B(kT_c)^5/k\theta.
\end{equation}
From Eq.~\eqref{eE2Tlimit} it follows that for $T\rightarrow T_c$, $\theta \propto B/E^2$ or
\begin{equation}
\label{eTeTc}
        T(E)\simeq T_c - Z^2 B(kT_c)^5/kE^2,
\end{equation}
that is, $T(E)$ \emph{monotonically increases with growing} $E$.

Returning to the main equation~\eqref{eP} of the present approach, it should be emphasized that it yields a \emph{single-valued and exact} simple relation between $T^\ast$ and $P^\ast$, while in the approach of Kunchur \cite{Kun02prl,Kni06prb} the calculation of the function $P^\ast(T^\ast)$ was possible only by \emph{numerical integration} of Eq.~\eqref{eHeatBalance}.

The next task is to derive an \emph{exact} formula for the critical temperature $T^\ast$ which does not depend on other critical parameters. To accomplish this, $E^\ast(T^\ast)$ can be calculated following two different ways. The first is obvious from considering Eq.~\eqref{eE2T}. It yields
\begin{equation}
\label{etildeE}
        \tilde E_0^\ast (T^\ast)= Z\sqrt{B} \left\{[(kT^\ast)^5 - (kT_0)^5)]/k(T_c - T^\ast)\right\}^{1/2}.
\end{equation}
The second way is exploiting, as previously, the condition $dj/dE = 0$, where $j = \sigma(T)E$. A simple calculation then yields $(dE/dT)_{E = E^\ast} = E^\ast/(T_c - T^\ast)$. Finally, taking into account that $d[\sigma(T)E^2]/dT = 5Ak(kT)^4$, one has
\begin{equation}
\label{etildeEshort}
        \tilde E_0^\ast (T^\ast) = Z\sqrt{5B} (kT^\ast)^2.
\end{equation}
A comparison of Eqs.~\eqref{etildeE} and~\eqref{etildeEshort} yields the following equation for $T^\ast$
\begin{equation}
\label{ekT5}
        (kT^\ast)^5 - (kT_0)^5 = 5(kT^\ast)^4 k (T_c - T^\ast),
\end{equation}
from which it follows that $T^\ast$ depends only on $T_0$ and $T_c$ and it does not depend on $A$, $B$, $R$, and $\sigma_n$. Equation~\eqref{ekT5} can be presented in another form
\begin{equation}
\label{e6T5}
        6T^{\ast5} - 5T_cT^{\ast4} - T_0^5 = 0.
\end{equation}
Finally, one obtains that
\begin{equation}
\label{eTast}
        T^\ast = (5/6)T_c + (T_0/6)(T_0/T^\ast)^4.
\end{equation}
From Eq.~\eqref{eTast} it follows that for $T_0 \leq T^\ast/2$ the dependence of $T^\ast$ on $T_0$ is very weak and $T^\ast\simeq(5/6)T_c$, i.\,e. $T^\ast$ depends only on $T_c$. It is interesting to note that the two-fluid approach also leads to $B$-independent $T^\ast$ as in Fig.~3 of Ref. \cite{Kni06prb}. It is curiously that if one applies Eq.~\eqref{eTast} for the estimation of $T^\ast$ in YBCO samples~\cite{Kun02prl,Kni06prb}, then one comes with essentially the same $T^\ast(T_0,T_c)$ dependence as obtained in Refs. \cite{Kun02prl,Kni06prb}, see e.\,g., Fig.~4 in Ref. \cite{Kni06prb}.

Now it is worth to return to the determination of the $(B,T^\ast)$-dependences of the other critical parameters, namely $\tilde j_0^\ast$, $\tilde v_0^\ast$, $\tilde \rho_0^\ast$, and $\tilde P^\ast$ in the presented approach for the flux-flow regime, using for that Eqs.~\eqref{eP}, \eqref{etildeE}, \eqref{etildeEshort}, and \eqref{ekT5}. The result is
\begin{equation}
\begin{array}{lll}
\label{e4}
        \tilde j_0^\ast = \sigma_n Z \sqrt{5}(kT^\ast)^2H_{c2}(T^\ast) /\sqrt{B},\\[1mm]
        \tilde v_0^\ast = c Z \sqrt{5}(kT^\ast)^2 /\sqrt{B},\\[1mm]
        \tilde \rho_0^\ast = \rho_n B/H_{c2}(T^\ast),\\[1mm]
        \tilde P_0^\ast = 5Ak (T_c - T^\ast)(kT^\ast)^4.
\end{array}
\end{equation}

A comparison of the critical parameters, obtained in the two-fluid approximation and given by Eqs.~\eqref{eTast},\eqref{e4} with the similar parameters in Ref. \cite{Kun02prl}, reveals that their $B$-dependences are identical. The merit of Eqs~\eqref{eTast},\eqref{e4} consists in that the $T^\ast$-dependent functions in these equations can be at once calculated using Eq~\eqref{eTast} for $T^\ast$. In other words, the presented two-fluid approach, based on a more simple heat balance equation~\eqref{eP}, allows one to derive the same results for the hot electron instability as obtained in~\cite{Kun02prl,Kni06prb} in a more direct and simple way \emph{without numerical integration} of Eq.~\eqref{eHeatBalance}. Introduction of pinning into the two-fluid approach follows the same way as it was discussed in Sec. \ref{SecInst} for the cosine and saw-tooth WPPs, i.\,e. using the function $2\tilde \mu = \tilde x = (j_c/\tilde j_0^\ast)^2$ with $\tilde j_0^\ast$ given by Eq.\eqref{e4}, as detailed next.

\subsection{Cosine potential}
Using Eqs.~\eqref{eCosine} and \eqref{eP}, in the presence of the cosine WPP the equation for $\tilde E^\ast$ reads
\begin{equation}
\label{eCosineLong}
        \left\{j_c^2 + [\sigma(T^\ast)\tilde E^\ast]^2  \right\}\tilde E^{\ast2}= A^2\left[(kT^\ast)^5 - (kT_0)^5\right]^2.
\end{equation}
Using Eq.~\eqref{etildeE}, Eq.~\eqref{eCosineLong} can be transformed into the previously derived Eq.~\eqref{eZ2} with $z\equiv (\tilde E^\ast/\tilde E_0^\ast)^2$ and $2\mu\equiv(j_c/\tilde j_0^\ast)^2$, where $\tilde E_0^\ast$ and $\tilde j_0^\ast$ are given by Eqs.~\eqref{etildeEshort} and \eqref{e4}. Here and in what follows the tilde denotes, as previously, the critical parameters derived in the two-fluid approach. The solution of Eq.~\eqref{eCosineLong} is given, as previously, by Eq.~\eqref{eZsolution}, the derivation of $(\tilde j^\ast/\tilde j_0)^2 = y$ repeats Eq.~\eqref{eY} and so on. Taking into account that all critical parameters given by Eqs.~\eqref{etildeEshort} and \eqref{e4} have the same $B$-dependences as in~\cite{Kun02prl,Kni06prb}, all the results obtained in Sec. \ref{SecInst} can be applied.

\subsection{Saw-tooth potential}
Using Eq.~\eqref{eSaw} for the CVC and Eq. \eqref{eP} for $P = Ej$ it is possible at once to obtain the equation for $\tilde E^\ast$ in the form
\begin{equation}
\label{eST1}
      \sigma(\tilde E^\ast)^2[1 + \sqrt{1 + (2j_c/\sigma\tilde E^\ast)^2}] =2 A [(kT^\ast)^5 -(kT_0)^5].
\end{equation}
A simple transformation of Eq. \eqref{eST1} (which releases the square root) with taking into account that $\sigma(T^\ast) A[(kT^\ast)^5 -(kT_0)^5] =(\tilde j_0^\ast)^2 $ leads to the previous result given by Eq. \eqref{eEE0}, namely
\begin{equation}
\label{eST2}
    \tilde E^\ast/\tilde E_0^\ast = 1/\sqrt{1 + \tilde x},
\end{equation}
 where $\tilde x = (j_c / \tilde j_0^\ast)^2$. The calculation of $\tilde j^\ast$ from Eqs. \eqref{eSaw} and \eqref{eST2} yields
\begin{equation}
\label{eST3}
       \tilde j^\ast/\tilde j_0^\ast = \sqrt{1 + \tilde x}.
\end{equation}
Equations \eqref{eST2} and \eqref{eST3} allow one to derive the results for $\tilde \rho^\ast$. $\tilde v^\ast$, and $\tilde P$ in the form calculated previously in Sec. \ref{SecInst}. In this way, all the results of Sec. \ref{SubsectsST} can be repeated.

\section{Discussion}
Before a comparison of the results obtained in this work with those of Kunchur \cite{Kun02prl,Kni06prb}, first it is suitable to briefly summarize the theoretical and experimental features of the hot-electron instability discussed in Refs. \cite{Kun02prl,Kni06prb} for epitaxial YBCO films at temperatures $T_0\leq T_c/2$. The heat balance equation~\eqref{eHeatBalance} is the basic equation of the considered electron overheating problem. It determines the nonlinear $T(E)$ behavior which is consistent with a nonlinear CVC, despite of the fact that the Bardeen-Stephen formula for the linear flux-flow conductivity with $T$-dependent $\sigma$ [due to $H_{c2}(T)$] is used. Unfortunately, Eq.~\eqref{eHeatBalance} allows one to find the $T(E)$ dependence and the CVC by numerical integration only. Using the $T(E)$ dependence it was also shown that $T^\ast = 76$\,K at $T_0 \ll T_c\approx 90$\,K and $T^\ast$ weakly depends on $T_0$ up to $T_0\approx 40$\,K~\cite{Kni06prb}. Finally, the $B$-dependences of the critical parameters without pinning (subscript ``0'') obtained in Refs. \cite{Kun02prl,Kni06prb} read
\begin{equation}
\label{eCriticalParam}
\begin{array}{lll}
        E_0^\ast \propto \sqrt{B},\qquad j_0^\ast \propto 1/\sqrt{B},\qquad v_0^\ast \propto 1/\sqrt{B},\\[2mm]
         \rho_0^\ast \propto B, \quad\qquad P_0^\ast\neq f(B).
\end{array}
\end{equation}
It should be recalled that the experimental results obtained for YBCO films~\cite{Kun02prl} were fitted, in neglect of pinning, to the predicted $B$-dependences by Eq.~\eqref{eCriticalParam} and the respective $(B,T_0)$-dependent CVCs \emph{without any adjustable parameters}.

Proceeding now to a brief discussion of the new results obtained in this work it is worth to begin with the description of the way of accounting for pinning in low-$T_c$ superconducting films. In fact, the introduction of pinning into the hot-electron instability problem here is phenomenological: Instead of the linear CVC $j = \sigma(T)E$ (at $T=const$) with $\sigma (T) = \sigma_n H_{c2}/B$ used by Kunchur \cite{Kun02prl,Kni06prb}, here the nonlinear CVC (at $T=const$) ``generated'' by the WPP and taken at $T=0$ has been used. Theoretically, it is possible to use the CVCs derived in Ref. \cite{Shk99etp,Shk08prb} at $T>0$ as well, however, in this work the consideration was limited to the two CVCs calculated at $T=0$ due to their simplicity. As the CVC curvature depends on the particular WPP used, two different simple WPPs have been used, namely a cosine WPP \cite{Shk08prb,Shk11prb} and a saw-tooth WPP \cite{Shk99etp,Shk06prb}, refer to Fig. \ref{fig1}. Both WPPs lead to the appearance of a new additional parameter in the CVC, $j_c$, the critical (depinning) current density which, in turn, depends on the WPP-specific parameters. The two CVC types addressed in this work can be realized in nanostructured superconducting films exhibiting a WPP \cite{Dob10sst,Dob11snm,Dob11pcs,Dob12njp,Dob16sst,Dob15apl,Dob15met}, see e.\,g. Ref.~\cite{Dob17pcs} for a review. The $\sigma(T)$ employed in Eqs.~\eqref{eCosine} and \eqref{eSaw} is the same as that used by Kunchur \cite{Kun02prl,Kni06prb}.

After the introduction of pinning, the task was to determine the $(j_c,B)$-dependences of the new critical parameters $E^\ast$, $j^\ast$, $v^\ast$, $\rho^\ast$ and $P^\ast$ for the CVC for the WPPs of both types. In Sec. \ref{SecInst}, formulae \eqref{eZsolution}--\eqref{eRast} for these critical parameters were obtained in terms of the dimensionless parameter $2\mu = (j_c/j_0^\ast)^2$ for the cosine WPP [see Eq.~\eqref{e2mu}], and $x\equiv 2\mu$ for the saw-tooth WPP. Then the problem of analyzing of the ($\mu,B$)-dependences of the aforementioned critical parameters was considered in two steps.
\begin{table*}[tbh!]
    \centering
\begin{footnotesize}
    \begin{tabular}{|c|c|c|c|c|}
\hline
    Two-fluid approach:                           &\multicolumn{4}{|c|}{$T^\ast\simeq(5/6)T_c$ for $T_0 \leq T^\ast/2$,   $\tilde \alpha=3\pi/4ecN(0)(kT_c)$      and   $\tilde \gamma=\sqrt{ \tilde P_0^\ast/\tilde \alpha}$} \\
\hline
  CVC: $j=\sigma E$, \newline
  $\sigma = \sigma_nH_{c2}(T)/B$                               & \multicolumn{2}{|c|}{cosine WPP: $ j = \sqrt{j_c^2 + \sigma ^2 E^2}$  for $j>j_c$}       & \multicolumn{2}{|c|}{saw-tooth WPP: $ j = (\sigma E/2)[1 + \sqrt{1+(2j_c/\sigma E)^2}$}\\
\hline
     clean limit $j_c=0$                             &  $ j_c=const $                & $j_c=j_c(B)$ Eq.~(17)                  &  $ j_c=const $             & $j_c=j_c(B)$ Eq.~(17)                              \\
\hline
    \multirow{1}{*}{$\tilde E_0^\ast =\tilde \gamma\tilde \alpha \sqrt{B}$} & \multicolumn{2}{|c|}{$\tilde E^\ast = \tilde E_0^\ast/\sqrt{\sqrt{1+\mu^2}+\mu}$}        & \multicolumn{2}{|c|}{$ \tilde E^\ast/= \tilde E_0^\ast /\sqrt{1 + \tilde x}$}\\
    \cline{2-5}
                                 & $ d\tilde E^\ast/dB >0$ Eq.(13)                 & $ d\tilde E^\ast/dB >0$ Eq.~(20)                   &  $ d{\tilde E^\ast}/dB >0$                 &  $ d\tilde E^\ast/dB >0$            \\
\hline
    \multirow{1}{*}{$\tilde j_0^\ast=\tilde \gamma/\sqrt{B}$} & \multicolumn{2}{|c|}{$\tilde j^\ast =\tilde j_0^\ast\sqrt{\sqrt{1+\mu^2}+\mu}$}        & \multicolumn{2}{|c|}{$ \tilde j^\ast=\tilde j_0^\ast \sqrt{1 + \tilde x}$}\\
    \cline{2-5}
                                 & $d\tilde j^\ast/dB<0$~Eq.~(15)                 &  $d\tilde j^\ast/dB<0$ Eq.~(23)                   &  $d\tilde j^\ast/dB<0$                 &  $d\tilde j^\ast/dB<0$            \\
\hline
    \multirow{1}{*}{$\tilde \rho_0^\ast =\tilde \alpha B$} & \multicolumn{2}{|c|}{$\tilde \rho^\ast =\tilde \rho^\ast_0/(\sqrt{1 + \mu^2} + \mu)$}        & \multicolumn{2}{|c|}{$\tilde \rho^\ast =\tilde \rho^\ast_0/(1+\tilde x)$}\\
    \cline{2-5}
                                 &  $d\tilde \rho^\ast/dB>0$~Eq.~(14)                  &  $d\tilde \rho^\ast/dB>0$ Eq.~(25)                   &  $d\tilde \rho^\ast/dB>0$                  &  $d\tilde \rho^\ast/dB>0$              \\
\hline
    \multirow{1}{*}{$ \tilde v_0^\ast = c\tilde \gamma\tilde \alpha/\sqrt{B}$, $d\tilde v^\ast/dB=- \tilde v_0^\ast/2B<0$} & \multicolumn{2}{|c|}{$\tilde v^\ast(B) = \tilde v_0^\ast /\sqrt{\sqrt{1+\mu^2} + \mu}$}        & \multicolumn{2}{|c|}{$\tilde v^\ast = \tilde v_0^\ast /\sqrt{1 + \tilde x}$}\\
    \cline{2-5}
                                 & $ d\tilde v^\ast/dB<0$ Eq.~(16)                  & $ d\tilde v^\ast/dB>0  or<0$ Eq.~(27)                  & $ d\tilde v^\ast/dB<0$                 & $ d\tilde v^\ast/dB>0  or<0$ Eq.~(32)            \\
\hline
   $ \tilde P_0^\ast =\tilde \gamma^2\tilde \alpha$                           & $\tilde P^\ast=\tilde P_0^\ast$                 & $\tilde P^\ast=\tilde P_0^\ast$                                  & $\tilde P^\ast=\tilde P_0^\ast$                                  & $\tilde P^\ast=\tilde P_0^\ast$                             \\
\hline
    \end{tabular}
\end{footnotesize}
    \caption{Summary of the results obtained in Sec. \ref{SecPow} within the \emph{two-fluid approach}. In the left column formulae for the critical parameters $\tilde E_0^\ast$, $\tilde j_0^\ast$, $\tilde \rho_0^\ast$, $ \tilde v_0^\ast$,  $\tilde P_0^\ast$ are derived in the clean limit, i.e. in absence of pinning ($j_c = 0$). They are presented in terms of the two parameters $\tilde \alpha$ and $\tilde \gamma$ calculated by Eqs.~\eqref{e4} taken at $T^\ast\simeq(5/6)T_c$ (see first the formulae for $\tilde \alpha$ and $\tilde \gamma$, $N(0)$ is the electron density of states of a metal film). In the second line of the table in the left column there is a CVC for the clean limit ($j_c = 0$), while in the second and third columns there are CVCs for the cosine and saw-tooth WPPs at $j > j_c$, respectively, which are both zero when $0< j < j_c$. The subsequent lines in the latter columns present the formulae for the pinning-dependent critical parameters $\tilde E^\ast$, $\tilde j^\ast$, $\tilde \rho^\ast$, $\tilde \rho^\ast$, $\tilde P^\ast$ and the behavior of their  $B$-derivatives in terms of  $2\mu=2\tilde \mu = \tilde x = (j_c/\tilde j_0^\ast)^2$, where $ j_c=const$ or $ j_c= j_c(B)$. }
 \label{table}
 \vspace{-5mm}
\end{table*}

First, it was supposed that $j_c$ is $B$-independent. Then, a direct inspection of Eqs.~\eqref{eZsolution}--\eqref{eRast} for the cosine WPP reveals that the critical parameters monotonically change with increasing $\mu$. Namely, at a fixed $j_0^\ast \propto 1/\sqrt{B}$, as $j_c$ increases, $E^\ast$, $\rho^\ast$, and $v^\ast \propto E^\ast$ decrease, $j^\ast$ increases, and $P^\ast$ does not depend on $j_c$. Analogous results have been derived for the saw-tooth WPP, see Eqs.~\eqref{eEE0}-\eqref{eRR0}. Then, taking into account Eq.~\eqref{eCriticalParam}, i.\,e. that $\mu\propto B$ at $j_c = const$, the $B$-dependence of the critical parameters and their $B$ derivatives for the cosine WPP [see Eqs.~\eqref{edEdB}-\eqref{edVdB}] have been analyzed. The main results of this analysis, which are similar for the cosine and the saw-tooth WPPs, can be summarized as follows: $E^\ast(B)$ monotonically increases with growing $B$ while $dE^\ast/dB > 0$ strongly decreases. The behavior of $\rho^\ast(B)$ upon $B$ is similar. $j^\ast(B)$ monotonically decreases with growing $B$ while its derivative $dj^\ast/dB <0$ strongly decreases. The critical velocity $v^\ast(B)$ and its derivative $dv^\ast/dB$ monotonically decrease with growing $B$. The power $P^\ast$ at the instability point is independent of $B$.

The second important step detailed in Sec.~\ref{SecInst} was to introduce a simple power-law dependence for $j_c(B)$ by Eq.~\eqref{eScaling}, because the previous assumption on the $B$-independence of $j_c$ is not realistic. In consequence of this, $\mu(B)\propto B^{1-2m}$ has a more complex $B$-dependence with $m\geq0$ that provides that $j_c(B)$ decreases with growing $B$ as observed in experiments. For $m=0$ one returns to the previous case with $j_c = const$ and at $m = 1/2$ a crossover appears from $\mu$ increasing with growing $B$ (for $0<m<1/2$) to $\mu$ decreasing in $B$ (for $m>1/2$). Turning to the influence of the $\mu(B)$ dependence on the $B$-behavior of the critical parameters and their $B$-derivatives, it has been derived that the $B$-derivatives of $E^\ast(B)$, $j^\ast(B)$, $\rho^\ast(B)$ at $m>0$ hold the same sign as for the case $j_c=const$. The $B$-behavior of $v^\ast(B)$ has been revealed to be quite different. Namely, $dv^\ast/dB$ changes its sign at $m>1$, i.\,e. $dv^\ast/dB>0$ at $B\rightarrow 0$. It should be noted that the main results of this analysis, as previously, are similar for both WPP types. Moreover, since usually $j_c(B)$ at $B\rightarrow0$ can be approximated by Eq.~\eqref{eScaling} with $m<1/2$, then \emph{ $dv^\ast/dB$ may exhibit a second sign change} at $B\ll B_{c2}$. This behavior is sometimes observed in experiments ~\cite{Gri09pcm,Leo10pcs,Gri10prb,Gri12apl,Gri11snm,Sil12njp}.

Finally, in Sec. \ref{SecPow} the simplest heat balance equation for electrons in low-$T_c$ superconducting films like Nb in the two-fluid approach has been considered. In this case the physics of quasiparticles overheating can be explained by a more simple heat balance equation than Eq.~\eqref{eHeatBalance}. The main features of this more simple approach were presented by BS in~\cite{Bez92pcs}, see Sections 1 and 2 therein. In using them it was supposed that $P(T,T_0)$ dependence can be approximated by the same expression, see Eq.~(18) in~\cite{Bez92pcs}, as for normal electrons at temperature $T$ near $T_c$ and this can be made within the framework of the two-fluid model of superconductivity~\cite{Tin04boo}. It was shown that $T^\ast$ appears near $T_c$ (but not too close to $T_c$ where the mechanism of the LO instability~\cite{Lar75etp} dominates). For the dissipated heat power $P$ flowing from the film to the substrate the heat balance equation has the form of Eq.~\eqref{eP} which is accurate to corrections of the order of $(\Delta/T)^2\ll 1$, where $\Delta(T)$ is the superconducting gap. Equation~\eqref{eP} describes the case when nonequilibrium phonons escape from the thin film without reabsorption by quasiparticles. The heating regime of the film in this limit is known as electron overheating~\cite{Bez92pcs}, termed so as one describes quasiparticles and phonons by different temperatures, $T$ and $T_0$, respectively. The main result of this section is Eq.~\eqref{eTast} from which follows that for $T_0 \leq T^\ast/2$ the dependence $T^\ast$ on $T_0$ is very weak and $T^\ast\simeq(5/6)T_c$, i.\,e. $T^\ast$ depends only on $T_c$.

A comparison of the critical parameters obtained in the clean limit within the framework of the two-fluid model [Eqs.~\eqref{eTast}--\eqref{e4}] with the respective parameters of Ref. \cite{Kun02prl}, see also Table \ref{table}, reveals that their $B$-dependences are identical. The merit of Eqs. \eqref{eTast} and \eqref{e4} consists in that the $T^\ast$-dependent functions in these equations can be at once calculated using Eq. \eqref{eTast} for $T^\ast$. In other words, the presented two-fluid approach based on a more simple heat balance equation~\eqref{eP} allows one to derive the same results for the hot electron instability as obtained by Kunchur \cite{Kun02prl,Kni06prb} in a more direct and simple way \emph{without numerical integration} of Eq.~\eqref{eHeatBalance}. Introduction of pinning into the two-fluid approach is done by the same way as discussed in Sec.\,\ref{SecInst} for the cosine and saw-tooth WPPs.

To draw parallels with the LO instability problem at $T \lesssim T_c$, it is worth emphasizing that a theoretical account for the pinning effect is possible in this case as well \cite{Shk17snd}. The introduction of pinning in Ref. \cite{Shk17snd} has been done in the same way as in this work, namely for a cosine WPP which can be realized in nanostructured low-$T_c$ superconducting films \cite{Dob10sst,Dob11pcs,Dob15apl,Dob15met,Dob12njp,Dob16sst}. The problem of pinning effects on the flux flow instability in Ref. \cite{Shk17snd} was at once considered relying upon the BS approach \cite{Bez92pcs} because one the LO instability corresponds to the limiting case of the BS instability at $B\ll B_T$, where $B_T$ is the quasiparticles overheating field introduced by BS in Ref. \cite{Bez92pcs}. In Ref. \cite{Shk17snd}, the heat balance equation in conjunction with the CVC extremum condition at the instability point has been augmented by a pinning strength parameter. A theoretical analysis \cite{Shk17snd} revealed that with increasing pinning strength at a fixed magnetic field value $E$ decreases, $j^\ast$ increases, while $P^\ast$ and $T\ast$ remain practically constant.

Lastly, turning to a comparison with experiment, it is worth noting that the presented account for pinning effects on the hot-electron vortex flow instability has recently allowed us to fit experimental data for the measured dependences $v^\ast(B)$ for epitaxial Nb films with different pinning types and its strength at $T = 0.4T_c$ \cite{Dob17arx} to the analytical expression \eqref{eVastB} derived here. In particular, we observed that the exponent $m \simeq 1$ in $j_c \propto 1 / B^m$ is larger in Nb films with stronger pinning (represented by ion-irradiated and nanopatterned films), while $m \simeq 0.5$ for as-grown films. In this way, we have been able to fit the observed crossover \cite{Dob17arx} from the monotonic decrease of $v^\ast(B)$ in the case of the as-grown films to the non-monotonic behavior of $v^\ast(B)$ for the films with stronger pinning.

\section{Conclusion}
To sum up, the proposed phenomenological approach for the introduction of pinning into the hot-electron instability problem has revealed the possibility for non-monotonicity of $v^\ast(B)$, as sometimes observed in experiments \cite{Gri09pcm,Leo10pcs,Gri10prb,Gri12apl,Gri11snm,Sil12njp,Dob17arx}. Addressing the experimental examination of the elaborated phenomenological theory, it should be pointed out that \emph{only two curves, namely the current-voltage characteristic and the $j_c(B)$ dependence have to be determined in experiment} thus allowing one to map the predicted results on the experimental data.

\section*{Acknowledgements}

The author thanks O. V. Dobrovolskiy for a proofreading of the manuscript. This work was supported through DFG project DO1511/3-1.
\vspace*{6mm}
%
\balance

\end{document}